\documentclass[prd,twocolumn,showpacs,amsmath,amssymb,superscriptaddress,floatfix,nofootinbib]{revtex4}
\usepackage{amsfonts}
\usepackage[colorlinks=true,citecolor=blue, linkcolor=blue,urlcolor=blue]{hyperref}
\usepackage{color}
\usepackage{graphicx,amsfonts,multirow}
\usepackage{amsmath}
\usepackage{ragged2e}

\newcommand{\ds}{\displaystyle}
\newcommand{\DS}[1]{/\!\!\!#1}
\definecolor{darkerblue}{rgb}{0,0,0.7}
\allowdisplaybreaks
\begin{document}

\title{$D \to P(\pi,K)$ helicity form factors within light-cone sum rule approach}

\author{Hai-Bing Fu}
\address{Institute of Particle Physics $\&$ Department of Physics, Guizhou Minzu University, Guiyang 550025, P.R. China}
\author{Wei Cheng\footnote{Corresponding author}}
\email{chengwei@itp.ac.cn}
\address{CAS Key Laboratory of Theoretical Physics, Institute of Theoretical Physics, Chinese Academy of Sciences,
Beijing 100190, China}
\address{Institute of Theoretical Physics, Chinese Academy of Sciences,  P. O. Box 2735, Beijing 100190, China}
\author{Rui-Yu Zhou}
\address{Department of Physics, Chongqing University, Chongqing 401331, P.R. China}
\author{Long Zeng}
\address{Institute of Particle Physics $\&$ Department of Physics, Guizhou Minzu University, Guiyang 550025, P.R. China}

\begin{abstract}
In this paper, the $D\to P(\pi, K)$ helicity form factors (HFFs) are studied by applying the QCD light-cone sum rule (LCSR) approach. The calculation accuracy is up to next-to-leading order (NLO) gluon radiation correction of twist-(2,3) distribution amplitude. The resultant HFFs at large recoil point are ${\cal P}_{t,0}^\pi(0) = 0.688^{+0.020}_{-0.024}$, ${\cal P}_{t,0}^K(0)=0.780^{+0.024}_{-0.029}$. In which, the contributions from three particles of the leading order (LO) are so small that can be safely neglected, and the maximal contribution of the NLO gluon radiation correction for ${\cal P}_{t,0}^{\pi,K}(0)$ is less than $3\%$. After extrapolating the LCSR predictions for these HFFs to whole $q^2$-region, we obtain the decay widths for semileptonic decay processes $D\to P\ell\nu_\ell$, which are consistent with BES-III collaboration predictions within errors. After considering the $D^{+}/D^{0}$-meson lifetime, we give the branching fractions of $D\to P\ell\nu_\ell$ with $\ell = e, \mu$, our predictions also agree with BES-III collaboration within errors, especially for $D\to \pi \ell\nu_\ell$ decay process. Finally, we present the forward-backward asymmetry ${\cal A}_{\rm FB}^\ell(q^2)$ and lepton convexity parameter ${\cal C}_F^\ell(q^2)$, and further calculate the mean value of these two observations $\langle{\cal A}_{\rm FB}^\ell\rangle$ and $\langle{\cal C}_F^\ell\rangle$, which may provide a way to test those HFFs in future experiments.
\end{abstract}
\date{\today}

\maketitle

\section{introduction}

Since the $D$-meson was first discovered by the Mark I detector at the Stanford Linear Accelerator Center (SLAC) in 1976, experimentalists have done a lot of research about its properties. For example, the decay constant is measured by Mark III detector at the SLAC~\cite{Adler:1987ty}, CLEO~ \cite{Eisenstein:2008aa, Artuso:2005ym, Bonvicini:2004gv} and BES~\cite{Ablikim:2004ry}, the corresponding branching fraction measurements can be obtained from Mark III~\cite{Adler:1989rw}, Belle~\cite{Widhalm:2006wz}, CLEO-c~\cite{Besson:2009}, BABAR~\cite{Lees:2015} and BES-III~\cite{ablikim:2016sqt, ablikim:2015, Ablikim:2015qgt}, and the corresponding semileptonic form factors are determined by BES-III~\cite{Ablikim:2018upe} and HPQCD Collaboration for lattice results~\cite{Koponen:2012di}, and so on. Recently, the BES-III collaboration update the new measurements for the $D$ meson semileptonic decay into pseudoscalar $\pi$ and $K$-meson processes, i.e., the absolute branching fractions are $\mathcal B(D^+ \to \bar K^0 e^+\nu_e)=8.60(6)(15)\times10^{-2}$, $\mathcal B(D^+ \to \pi^0 e^+\nu_e) = 3.63(8)(5)\times10^{-3}$~ \cite{ablikim:2017lks}. A brief review of the earlier work and current experimental status of $D$-meson decays can refer to Ref.~\cite{Amhis:2016xyh}.

The $D$-meson is the lightest mesons containing a single charm quark (antiquark) and has an abundance of decay channels.
Among them, the heavy-to-light decay contains a lot of information for the dynamics of weak and strong interaction~\cite{Faustov:2019mqr}, which may provide a platform to test the standard model more accurately and seek new physics (NP) beyond the SM. For example, the semileptonic decay $D\to P\ell\nu_\ell$ processes can be used to extract the Cabibbo-Kobayashi-Maskawa (CKM) matrix elements, which is an important part of SM, due to the associated decay branching fractions are proportional to the CKM matrix elements~\cite{Charles:2004jd,Zhou:2019crd}. Meanwhile, a single phase from the CKM quark mixing matrix will dominate the CP violation phenomena~\cite{Deshpande:1997rr}, which is relevant to NP. Thus, the $D$-meson semileptonic decay is widely studied by various theoretic methods.

The usual strategy for investigating the $D$-meson semileptonic decay is to express the decay process as non-perturbative hadronic matrix element with different $\gamma$-structures, and factorize it into Lorentz invariant transition form factors (TFFs) by employing covariant decomposition, and then study those TFFs by various theoretic methods, such as LCSRs~\cite{svz, Ball:1997rj, Ball:2004rg, Huang:2008zg, AKhodjamirian:2010, Fu:2014cna, Cheng:2017bzz, Ahmady:2013cga, Straub:2015ica,Wang:2015vgv}, the transverse-momentum-dependent (TMD) factorization approach~\cite{Li:2012nk,Li:2012md}, the lattice QCD~\cite{Lattice96:1, DelDebbio:1997nu, Lattice04, Horgan:2013hoa, Horgan:2013pva, Agadjanov:2016fbd} and the perturbative QCD~\cite{Kurimoto:2001zj, Chen:2002bq, Kurimoto:2002sb, Keum:2004is, Fan:2013qz, Lu:2018cfc, Lu:2018obb, Gao:2019lta} etc. As an alternative strategy, with other processes remain unchanged, only the non-perturbative hadronic matrix elements are projected as the Lorentz invariant HFFs through the covariant helicity factor. The main difference with usual TFFs is lying in different projection methods, which will bring some unique advantages to HFFs and may get better physical predictions. These can refer to Refs.~\cite{Bharucha:2010im,Cheng:2018ouz}. For example, the method is used in dealing with the $B \to \rho$ decays in our previous work~\cite{Cheng:2018ouz}, and the resultant differential decay width is in good agreement with BABAR experiment. In this paper, we will attempt to use HFFs within the framework of LCSR to study the $D \to P(\pi,K)$ decays. Note that, in order to get more accuracy LCSR results, our calculations for the $D \to P(\pi,K)$ HFFs, ${\cal P}_{0,t}^P(q^2)$ contain both LO and NLO contributions. For the LO contributions, two and three particles parts are contined up to twist-4 light-cone distribution amplitudes (LCDAs), while for the NLO contributions, one-loop gluon radiation correction is contained for the twist-(2,3) LCDAs.

The remaining parts of the paper are organized as follows. In Sec.~\ref{sec:2}, we give a brief introduce for the definition of HFFs and provides the full LCSR expression for ${\cal P}^P_{0,t}(q^2)$. In Sec.~\ref{sec:Numerical}, we first discuss the hadron input parameters for HFFs and extrapolate those HFFs to the whole $q^2$ region by employing SSE. Then, we compute the differential decay width and branching fraction for $D\to P(\pi,K)$ decays from the HFFs, We compare our results with available experimental other theoretical results. Finally, the conclusion is given in Sec.~\ref{sec:summary}.

\section{Calculation Technology}\label{sec:2}

The short distance hadronic matrix elements related to the pseudoscalar $D$-meson semileptonic decays can be projected as the relevant HFFs via the off-shell $W$-boson polarization vectors with 4-momentum $q^\mu=(q^0,0,0,-|\vec q\,|)$~\cite{Bharucha:2010im},
\begin{align}
\label{eq:AVdef}
{\cal P}_{\sigma}(q^2 )& =\sqrt{\frac{q^2}{\lambda}} \,
  {\varepsilon_\sigma ^{*\mu}(q)}
\, \langle P(k)|\bar q \, \gamma_\mu \, c |D(p)\rangle \,.
\end{align}
where the standard kinematic function $\lambda = (t_- - q^2)(t_+ - q^2)$ with $t_\pm = (m_D\pm m_P)^2$. The polarization vectors $\varepsilon_\sigma ^{*\mu}(q)$ represent transverse ($\sigma=\pm$), longitudinal ($\sigma=0$) or time-like ($\sigma=t$) component, more specifically,
\begin{eqnarray}
 \varepsilon _ \pm ^\mu(q) &=&  \mp \frac{1}{\sqrt 2 } (0,1, \mp i,0) ,\\
\qquad
 \varepsilon _0^\mu  (q) &=&   \frac{1}{\sqrt {q^2 } } (|\vec q| ,0,0, - q^0 ) ,\\
\cr
 \varepsilon _t^\mu  (q) &=& \frac{1}{\sqrt {q^2 } }  q^\mu .
\end{eqnarray}

In order to derive the full analytical LCSRs expression for those HFFs, we take the following two-point correlation function as a start point,
\begin{align}
\Pi_\sigma(p,q) &= i \sqrt{\frac{q^2}{\lambda}} {\varepsilon_\sigma^{*\mu}(q)}
\nonumber\\
&\times\int d^4 x e^{iq\cdot x}\langle P(k)|T\{j_V^\mu(x),j_D^\dag (0)\}|0\rangle,\label{correlators}
\end{align}
where the current $j_V^\mu(x) = \bar {q}(x){\gamma _\mu }  c(x)$ and $j_D^\dag (0)=\bar c(0)i \gamma_5 u(0)$, which has the same quantum state as the pseudoscalar $D$-meson with $J^{P}=0^-$. $T$ stands for the product of the current operator.

Following the basic procedure of LCSR, the correlation function can be treated by inserting complete intermediate states with the same quantum numbers as the current operator $\bar c i \gamma_5 u$ in the time-like $q^2$-region. After isolating the pole term of the lowest pseudoscalar $D$-meson, one can reach the following expression,
\begin{align}
\Pi_\sigma ^H(p,q)&= \sqrt{\frac{q^2}{\lambda}} {\varepsilon_\sigma^{*\mu}(q)}\bigg[ \frac{ \langle P|\bar{q}\gamma _\mu c|D\rangle \langle D|\overline{c}i \gamma_5 u|0\rangle} {m_D^2-(p+q)^2} \nonumber \\[1ex]
&+\sum\limits_{\rm H}\frac{\langle P|\bar{q}\gamma _\mu c|D^{\rm H}\rangle \langle D^{\rm H}|\bar{q}i\gamma
_5 u|0\rangle}{m_{D^{\rm H}}^2-(p+q)^2}\bigg],
\end{align}
where $\langle D|\bar{c}i\gamma_5 u|0\rangle=m_D^2f_D/m_c$. Employing dispersion integrations to replace the contributions of higher resonances and continuum states, the hadronic representation of the correlator $\Pi^H_\sigma$ can be read off
\begin{align}
&\Pi^H_\sigma(q^2,(p+q)^2)=\frac{ m_D^2 f_D}{m_c [m_D^2-(p+q)^2]}{\cal P}_{\sigma}(q^2 )\nonumber\\
& \qquad + \int _{s_0}^{\infty}\frac {\rho^{\rm H}_\sigma(s)}{s-(p+q)^2}ds +{\rm subtractions},
\end{align}
where $s_0$ is the effective threshold parameter, and the spectral densities $\rho^{\rm H}_\sigma(s)$ can be approximated by employing the quark-hadron duality ansatz
\begin{equation}
\rho ^{\rm H}_\sigma(s)=\rho_\sigma ^{\rm QCD}(s)\theta (s-s_0).
\end{equation}

On the other hand, in the space-like $q^2$-region, i.e. $(p+q)^2-m_c^2\ll 0$ and $q^2\ll m_c^2-{\cal O}(1{\rm GeV ^2})$, the correlation function can be pre-processing by contracting the $c$-quark operates to a free propagator,
\begin{align}
&\langle0|c_\alpha^i(x)\bar c_\beta^j(0)|0\rangle = -i\int \frac{d^4k}{(2\pi)^4}e^{-ik\cdot x}\bigg\{\delta^{ij}\frac{\DS k + m_c}{m_c^2-k^2}\nonumber \\
&\quad +g_s\int_0^1 dv G^{\mu\nu\alpha}(vx)\left(\frac{\lambda}{2}\right)^{ij}\bigg[\frac{\DS k+m_c}{2(m_c^2 - k^2)^2}\sigma_{\mu\nu} \nonumber \\
&\quad+ \frac1{m_c^2-k^2}vx_\mu\gamma_\nu\bigg]\bigg\}_{\alpha\beta}.
\end{align}
where the nonlocal matrix elements are convoluted with the meson LCDAs with growing twist. The matrix elements can be expanded up to twist-4 LCDAs as follows ~\cite{Ball:2004ye}:

\begin{align}
&\langle P(p)|\bar q_1(x)i\gamma_5 u(0)|0\rangle = \frac{f_P m_P^2}{m_u + m_{q_1}} \int_0^1 du e^{iup\cdot x}\phi_{2;P}(u),\\
&\langle P(p)|\bar q_1(x)\gamma_\mu \gamma_5 u(0)|0\rangle = - i p_\mu f_P \int_0^1 du e^{iup\cdot x}\bigg[\phi_{3;P}^p(u)
\nonumber\\
&\quad+x^2\psi_{4;P}(u)\bigg]+f_P(x_\mu-\frac{x^2p_\mu}{p \cdot x})\int_0^1 du e^{iup\cdot x}\phi_{4;P}(u),\\
&\langle P(p)|\bar q_1(x)\sigma_{\mu \nu} \gamma_5 u(0)|0\rangle =  i(p_\mu x_\nu - p_\nu x_\mu) \frac{f_P m_P^2}{6(m_u + m_{q_1})}
\nonumber\\
&\quad\times\int_0^1 du e^{iup\cdot x}\phi_{3;P}^\sigma(u),
\\ \nonumber\\
&\langle P(p)|\bar q_1(x)\sigma_{\mu\nu}\gamma_5 g_s G_{\alpha\beta}(vx)u(-x)|0\rangle = i f_{3P} (p_\alpha p_\mu g_{\nu\beta}^\bot
\nonumber\\
&\quad- p_\alpha p_\nu g_{\mu\beta}^\bot - (\alpha \leftrightarrow \beta))\Phi_{3;P}(v, p\cdot x),
\end{align}
\begin{align}
&\langle P(p) | \bar q_1(x)\gamma_\mu\gamma_5
gG_{\alpha\beta}(vx)u(-x)|0\rangle
\nonumber\\
&\qquad =  p_\mu (p_\alpha x_\beta- p_\beta x_\alpha) \frac{f_{P}}{p\cdot x}
\Phi_{4;P}(v,p\cdot x) \cr
&\qquad + (p_\beta g_{\alpha\mu}^\perp -p_\alpha g_{\beta\mu}^\perp) f_{P}\Psi_{4;P}(v,p\cdot x)
\\ \nonumber\\
&\langle  P(p) | \bar q_1(x)\gamma_\mu i
g\widetilde{G}_{\alpha\beta}(vx)u(-x)|0\rangle  \nonumber\\
&\qquad =  p_\mu (p_\alpha x_\beta - p_\beta x_\alpha) \frac{f_P}{p\cdot x}
\widetilde\Phi_{4;P}(v,p\cdot x)  \cr
&\qquad+ (p_\beta g_{\alpha\mu}^\perp -
p_\alpha g_{\beta\mu}^\perp) f_{P} \widetilde\Psi_{4;P}(v,p\cdot x),
\label{4.16}
\end{align}
where the $P$ stands for $\pi,K$-meson with $q_1=d,s$ quark, and we have set
\begin{align}
& g_{\mu\nu}^\bot = g_{\mu\nu} - \frac{p_\mu x_\nu + p_\nu x_\mu}{p\cdot x}, \cr
& K(v,p\cdot x) = \int_0^1 d\alpha_1 d\alpha_2 d\alpha_3 \delta(1-\alpha_1-\alpha_2-\alpha_3) K(\alpha_i).\nonumber
\end{align}
Here $K(\alpha_i)$ stands for the twist-3 or twist-4 DA $\Phi_{3;P}(\alpha_i)$, $\Phi_{4;P}(\alpha_i)$, $\Psi_{4;P}(\alpha_i)$, $\widetilde\Psi_{4;P}(\alpha_i)$ and $\widetilde\Psi_{4;P}(\alpha_i)$. Furthermore, by replacing the non-local matrix elements and using dispersion integration to carry out the subtraction procedure of the continuum spectrum, the QCD representation can be obtained.

After equating the two type representation of correlator and subtracting the contribution from higher resonances and continuum states, one further carries out Borel transformation for them, i.e. replacing the variable $(p + q)^2$ with the Borel parameter $M^2$ and exponentiating the denominators, which removes the subtraction term in the dispersion relation and exponentially suppresses the contributions from unknown excited resonances. To obtain higher calculation accuracy, we calculate the NLO gluon radiation correction of twist-(2,3) distribution amplitude for the $D \to P$, using the same method as the treatment $B\to \pi$ process~\cite{Duplancic:2008ix}. The one-loop corrections may lead to amplitude divergence, which can be separated as UV divergence and infrared collinear divergence. The UV divergence will be canceled by the mass renormalization of heavy quark $c$ and infrared collinear divergence term can be absorbed by the evolution of the wave function of the meson\cite{Li:2015cta}. Thus, the LCSR for the $D \to P$ HFFs is finally obtained:
\begin{widetext}
\begin{align}
{\cal P}^P_0(q^2)&= \frac{f_P m_c}{f_D m_D^2} \int_0^1 d u e^{(m_D^2 - s)/M^2} \bigg\{ m_c \bigg[ \frac{1}{2u} \Theta (c(u,s_0))\phi_{3;P}^p(u) - \frac{2m_c^2}{u^3M^4} \widetilde{\widetilde\Theta}(c(u,s_0))\psi_{4;P}(u) + \frac{1}{u M^2} \widetilde\Theta (c(u,s_0)) \phi_{4;P}(u)
\nonumber\\
&+ \frac{2 m_c^2}{u^3 M^4} \widetilde{\widetilde\Theta}(c(u,s_0)) \Phi_{4I;P}(u) \bigg]
+\frac{m_P ^2}{2(m_u + m_{q_1})}\bigg[\Theta (c(u,s_0)) \phi_{2;P}(u) + \bigg( \frac{1}{3u} \Theta (c(u,s_0)) + \frac{m_c^2 + q^2}{6u^2M^2} \widetilde\Theta (c(u,s_0))\bigg)
\nonumber\\
&\times \phi_{3;P}^\sigma(u) \bigg] - \frac{f_{3\pi}}{f_P }\frac{I_{3\pi}(u)}{u}- \frac{m_c}{2(m_c^2-q^2)}I_{4\pi}(u) \bigg\} + \frac{\alpha_s C_F}{8\pi m_D^2 f_D}F_1(q^2,M^2,s_0),
\label{eq:Av0}\\
\nonumber\\
{\cal P}^P_t (q^2)&  = \frac{f_P m_c}{\sqrt \lambda f_D m_D^2 }\int_0^1 d u e^{(m_D^2 - s)/M^2} \bigg\{{\cal Q}_+ m_c \bigg[ \frac{1}{2u} \Theta (c(u,s_0))\phi_{3;P}^p(u) - \frac{2m_c^2}{u^3M^4} \widetilde{\widetilde\Theta}(c(u,s_0))\psi_{4I;P}(u) + \frac{1}{u M^2} \widetilde\Theta (c(u,s_0))\nonumber\\
&\times \phi_{4;P}(u)+ \frac{2 m_c^2}{u^3 M^4} \widetilde{\widetilde\Theta}(c(u,s_0))
\Phi_{4;P}(u) \bigg]+\frac{{\cal Q}_+m_P ^2}{2(m_u + m_{q_1})}\bigg[\Theta (c(u,s_0))\phi_{2;P}(u) + \bigg( \frac{1}{3u} \Theta (c(u,s_0)) + \frac{m_c^2 + q^2}{6u^2M^2}
\nonumber\\
& \times \widetilde\Theta (c(u,s_0))\bigg)\phi_{3;P}^\sigma(u) \bigg]+ q^2 \bigg[\frac{4 m_c}{u^2 M^2 } \widetilde\Theta (c(u,s_0))\phi_{4;P}(u) + \frac{2 m_P ^2}{u(m_u + m_{q_1})}\Theta (c(u,s_0))\phi _{2;P}(u)\bigg]\bigg\}+\frac{m_D^2-m_P^2}{\sqrt{\lambda}}
\nonumber\\
&\times\int_0^1 du e^{(m_D^2 - s)/M^2}\frac{{\cal Q}_-}{m_D^2 - m_P^2}\bigg[-\frac{m_cf_{3P}}{u m_D^2 f_D}I_{3;P}(u) - \frac{m_c^2 f_P}{2m_D^2 f_D(m_c^2-q^2)}I_{4;P}(u)\bigg] +\frac{\alpha_s C_F}{4\pi}~\frac{m_D^2-m_P^2}{2m_D^2 f_D\sqrt{\lambda}}
\nonumber\\
&\times \bigg[F_1(q^2,M^2,s_0) + \frac{q^2}{m_D^2-m_P^2}(\widetilde{F}_1(q^2,M^2,s_0) - F_1(q^2,M^2,s_0))\bigg]
\label{eq:Avt}
\end{align}
\end{widetext}
where $P$ stands for $\pi,K$-meson, ${\cal Q}_\pm = q^2 \pm (m_D^2 - m_P^2)$. The short-hand notations introduced for the integrals over three-particle DA's are:
\begin{align}
I_{3;P}(u)& =\frac{d}{du}\Bigg[\int\limits_0^u d\alpha_1\!\!\!\int\limits_{\frac{(u-\alpha_1)}{(1-\alpha_1)}}^1\!\!\!\!\! dv \Phi_{3;P}(\alpha_i)
\Bigg|_{\begin{array}{l}
\alpha_2=1-\alpha_1-\alpha_3,\\
\alpha_3=(u-\alpha_1)/v
\end{array}}\Bigg]\,,
\nonumber\\
\nonumber\\
I_{4;P}(u)&=\frac{d}{du}\Bigg\{\int\limits_0^u\! d\alpha_1\!\!\!
\int\limits_{\frac{(u-\alpha_1)}{(1-\alpha_1)}}^1\!\!\!\!\! \frac{dv}{v} \,\,
\Bigg[2\Psi_{4;P}(\alpha_i)-\Phi_{4;P}(\alpha_i)
\nonumber \\
&  +2\widetilde{\Psi}_{4;P}(\alpha_i)-\widetilde{\Phi}_{4;P}(\alpha_i)\Bigg]
\Bigg|_{\begin{array}{l}
\alpha_2=1-\alpha_1-\alpha_3,\\
\alpha_3=(u-\alpha_1)/v
\end{array}
}\Bigg\}\,.
\label{eq:fplusBpiLCSR3part}
\end{align}
The NLO terms in Eqs.~\eqref{eq:Av0} and \eqref{eq:Avt} have the form of the dispersion relation:

\begin{align}
&F_1(q^2,M^2,s_0) = \frac{1}{\pi}\int\limits_{m_c^2}^{s_0}
ds e^{(m_D^2-s)/M^2}\,\mbox{Im} F_1(q^2,s)
\nonumber\\
&~ = \frac{f_P}{\pi} \int\limits_{m_c^2}^{s_0}ds e^{(m_D^2-s)/M^2}
\int_0^1 du\bigg \{ {\rm Im} T_1(q^2,s,u)\phi_{2;P}(u)\nonumber\\
&~ +  {\rm Im} T_1^p(q^2,s,u)\phi_{3;P}^p(u) + {\rm Im} T_1^\sigma(q^2,s,u)\phi_{3;P}^\sigma(u)\bigg \},
\label{eq:Imconvol}
\end{align}
where the expressions of the imaginary parts of the amplitudes can refer to $B\to \pi$ process~\cite{Li:2015cta}. The NLO amplitudes $\widetilde{F}_1(q^2,s,u)$ have the same expression as $T_1^{(p,\sigma)} (q^2,s,u) \to \widetilde T_1^{(p,\sigma)} (q^2,s,u)$. At zero momentum transfer, the additional relation ${\cal P}_t^P(0)={\cal P}_0^P(0)$ holds, which can be confirmed not only from the above the HFFs but also from Table~\ref{HFF uncertainties}, \ref{tab:LONLO} and Fig.~\ref{HFF:P0t}.

\section{Numerical Analysis}\label{sec:Numerical}
In order to do the numerical calculation, we take the $D$, $K$ and $\pi$-meson decay constants, $f_D=0.2037\pm0.0047\pm0.0006~{\rm GeV}$\cite{Patrignani:2016xqp}, $f_K=0.1555~{\rm GeV}$\cite{Khodjamirian:2009ys} and $f_\pi=0.1304~{\rm GeV}$\cite{Khodjamirian:2009ys}, the $c$-quark pole mass $m_c=1.28\pm0.03~{\rm GeV}$\cite{Tanabashi:2018oca}, the $D$, $K$ and $\pi$-meson mass ${m_D} = 1.865~{\rm GeV}$, ${m_{K}} = 0.494~{\rm GeV}$ and ${m_{\pi}} = 0.140~{\rm GeV}$\cite{Tanabashi:2018oca}. The factorization scale $\mu$ is set as the typical momentum transfer of $D \to K (\pi)$, i.e. $\mu_{\rm IR} \simeq  (m_{D}^2-m_c^2)^{1/2} \sim 1.3~{\rm GeV}$\cite{Fu:2018yin}.

\subsection{Distribution Amplitudes}
With regard to the hadron input about LCDAs, we will shortly discuss the twist-2,3,4 LCDAs. For the leading twist-2 LCDAs, we adopt a standard approach to do the calculation, i.e. the conformal expansion\cite{Ball:2004ye}:
\begin{equation}\label{eq:confexp}
\phi_{2;P}(u,\mu^2) = 6 u \bar u \left[ 1 + \sum\limits_{n=1}^\infty a^P_{n}(\mu^2) C_{n}^{3/2}(\xi)\right],
\end{equation}
where $\xi=2u-1$, $\mu$ is the factorization scale, the superscript $P$ represent the $\pi$ and $K$-meson, and $a^P_{n}$ is the (non-perturbative) Gegenbauer moments, which is usually up to the first two terms $a^P_{1,2}$ accuracy due to the suppression of large $n$ physical amplitudes of the Gegenbauer polynomials oscillate rapidly.

The two-particle twist-3 and twist-4 LCDAs $\phi_{3;P}^p(u)$, $\phi_{3;P}^{\sigma}(u)$, $\phi_{4;P}(u)$, $\psi_{4;P}(u)$ and $\Phi_{4I;P}(u)$ are defined as follows\cite{Ball:2006wn}:
\begin{align}
\phi_{3;P}^p(u)& = 1+(30\eta_3^P - \frac{5}{2}\rho_{\pi}^2)C^{1/2}_2(\zeta)+(-3\eta_3^P \omega_3^P
\nonumber\\
& -\frac{27}{20}\rho_{\pi}^2-\frac{81}{10}\rho_{\pi}^2a_2^P)C^{1/2}_4(\zeta), \\
\phi_{3;P}^{\sigma}(u)&= 6u(1-u)(1+(5\eta_3^P - \frac{1}{2}\eta_3^P\omega_3^P - \frac{7}{20}\rho_{\pi}^2
\nonumber\\
&-\frac{3}{5}\rho_{\pi}^2a_2^P) C^{3/2}_2(\zeta),\\
\phi_{4;P}(u)&=-m_P^2\int_0^u B(v)dv, \\
\psi_{4;P}(u)&=\frac{1}{4}m_P^2 A(u) - \int_0^u \phi_{4;P}(v)dv,\\
\Phi_{4I;P}(u)&=\int_0^u \phi_{4;P}(v)dv.
\end{align}
where
\begin{table}[t]
\centering
\caption{One-loop anomalous dimensions of hadronic parameters in DAs.}
\begin{tabular}{c c}
\hline\\[-10pt]
~~~~~$\gamma_{a_n}$~~~~~&~~~~~$C_F\left(1-\dfrac{2}{(n+1)(n+2)}-\ds\sum_{m=2}^{n+1}\dfrac{1}{m}\right)$~~~~~ \\[10pt]
\hline \\[-10pt]
$\gamma_{\eta_3}$   & $\dfrac{16}{3}C_F+C_A$             \\[2ex]
$\gamma_{\eta_4}$   & $\dfrac{8}{3}C_F$                  \\[2ex]
$\gamma_{\omega_3}$ & $-\dfrac{25}{6}C_F+\dfrac{7}{3}C_A $\\[2ex]
$\gamma_{\omega_4}$ & $-\dfrac{8}{3}C_F+\dfrac{10}{3}C_A$ \\[1.2ex]
\hline
\end{tabular}

\label{tab:anomalous}
\end{table}
\begin{align}
A(u) & =  6u\bar u \bigg[ \frac{16}{15} + \frac{24}{35}  a_2^P + 20 \eta_3^P + \frac{20}{9}\eta_4^P  + \bigg( -\frac{1}{15}   \nonumber\\
& + \frac{1}{16} - \frac{7}{27} \eta_3^P \omega_3^P - \frac{10}{27}\eta_4^P \bigg) C_2^{3/2}(\xi) + \bigg( -\frac{11}{210}  \nonumber\\
& \times a_2^P - \frac{4}{135} \eta_3^P \omega_3^P \bigg)
  C_4^{3/2}(\xi) \bigg]+ \bigg(-\frac{18}{5} a_2^P + 21  \nonumber\\
& \times \eta_4 ^P\omega_4^P \bigg) \bigg[ 2u^3 (10-15 u + 6 u^2)\ln u + 2\bar u^3 (10  \nonumber\\
& -15\bar u + 6 \bar u^2) \ln\bar u + u \bar u (2 + 13u\bar u)\bigg],
\\
\nonumber\\
B(u)&= 1+[1+\frac{18}{7}a_2^P + 60\eta_3^P + \frac{20}{3}\eta_4^P]C^{1/2}_2(\zeta)
\nonumber\\
&+[-\frac{9}{28}a_2^P - 6\eta_3^P\omega_3^P]C^{1/2}_4(\zeta).
\end{align}
As for LO contributions, they do not mix under renormalization in QCD theory, so that the scaling up to $\mu_{\text{IR}}$ is given by
\begin{align}
c_i(\mu_{\text{IR}}) = {\cal L}^{\gamma_{c_i}/\beta_0} c_i(\mu_0) ,\label{RF}
\end{align}
where ${\cal L} = \alpha_s(\mu_{\text{IR}})/\alpha_s(\mu_0)$, $\beta_0=11-2/3N_f$, and the one-loop anomalous dimensions $\gamma_{c_i}$ are given in Table~\ref{tab:anomalous}~\cite{Ball:2006wn}. Given the initial scale $\mu_0$ values of the hadronic parameters~\cite{Ball:2006wn}, employing the renormalization function Eq.\eqref{RF}, one can be obtained the corresponding values at the typical scale $\mu = \mu_{\rm IR}$, our results are listed in Table~\ref{tab:DA-num}.

\begin{table}
\caption{Hadronic parameters for the $\pi$ and $K$ DAs. In which the $\mu_0 = 1~{\rm GeV}$ and $\mu_{\rm IR} = 1.3~{\rm GeV}$.}\label{tab:DA-num}
$$\begin{array}{cccccc}
\hline
K      & \mu_0     & \mu_{\rm IR} ~~~~& ~~~~\pi~~~          & \mu_0    & \mu_{\rm IR} \\ \hline
a_1^K  & 0.06(3)   & 0.06(3)      & a_1^\pi      & 0        & 0            \\
a_2^K  & 0.25(15)  & 0.25(15)     & a_2^\pi      & 0.25(15) & 0.25(15) \\
\eta_3^K   & 0.015 & 0.011        & \eta_3^\pi   & 0.015    & 0.011 \\
\eta_4^K   & 0.6   & 0.542        & \eta_4^\pi   & 10       & 9.037 \\
\omega_3^K & -3    & -2.879       & \omega_3^\pi & -3       & -2.879 \\
\omega_4^K & 0.2   & 0.166        & \omega_4^\pi & 0.2      & 0.166 \\\hline
\end{array}$$
\end{table}

\subsection{Fixing effective threshold $s_0$ and Borel parameter $M^2$}

There are two ``internal'' parameters for the HFFs. One is an effective threshold parameter $s_0$, which is an output of the continuum subtraction procedure. The other one is Borel windows $M^2$, which comes from the Borel transformation to sum rules to suppress contributions of the higher resonances and continuum state~\cite{Ball:2004rg}. For the former, we take the effective threshold $s_0({\cal P}_{0;t}^\pi)=12(1)~\rm{GeV^2}$ and $s_0({\cal P}_{0;t}^K)=21(1)~\rm{GeV^2}$. For the Borel windows $M^2$, we set the continuum contribution to less than $25\%$ of the total LCSR to obtain the upper limit of $M^2$, i.e.
\begin{eqnarray}
\dfrac{\ds\int_{s_0}^\propto ds \rho^{\rm tot}(s)e^{-s/M^2}} {\ds\int_{m_c^2}^\propto ds\rho^{\rm tot}(s)e^{-s/M^2}} \le 25\%. \label{con65}
\end{eqnarray}
As the stability of $M^2$ is an important requirement of the sum rule calculation and the unified criteria, we will adopt a stability Borel windows $M^2$ to obtain the lower limit of $M^2$, i.e., the HFFs to be changed less than $2\%$ within the Borel window. The determined Borel parameter are $M^2({\cal P}_{0;t}^\pi) = 29.15\pm2.55~{\rm GeV}^2$ and $M^2({\cal P}_{0;t}^K) = 151\pm42.5~{\rm GeV}^2$.

After determined the LCSRs parameters, we give the HFFs ${\cal P}_{0;t}^{P}(q^2)$ at large recoil point $q^2\rightsquigarrow 0$ changed with various input parameters in Table~\ref{HFF uncertainties}, where the uncertainties are from the $D$-meson decay constant $f_D$, the $c$-quark pole mass $m_c$, the Borel parameter $M^2$ and the continuum threshold $s_0$. It shows that the main errors of those HFFs come from the parameters LCDAs and the effective threshold parameter $s_0$. Furthermore, the central value of LO and NLO contributions for the ${\cal P}_{0;t}^{P}(0)$ are shown in Table~\ref{tab:LONLO}. In which the maximal NLO contributions of ${\cal P}_{0;t}^P (0)$ are no more than $3\%$, which means our HFFs maintain high accuracy. Thus, it is reliable to use the HFFs for analysis of the $D \to P$ decays.

\begin{table}[t]
\caption{Uncertainties of the LCSR predictions on the HFFs ${\cal P}_{0;t}^{P}(0)$ caused by the errors of the input parameters. }\label{HFF uncertainties}
\centering
\begin{tabular}{c c c c c c c}
\hline	
~~ &~~${\rm {CV}}$~~ & ~~$\Delta{\rm DA}$~~ & ~~$\Delta {s_0}$~~ & ~~$\Delta {M^2}$~~ & ~~$\Delta ({m_c};{f_D})$~~\\
\hline
${\cal P}_{0;t}^\pi (0) $ & $0.688$ & $^{+0.004}_{-0.007}$ & $^{+0.017}_{-0.020}$& $^{+0.007}_{-0.008}$ & $^{+0.008}_{-0.009}$ \\
${\cal P}_{0;t}^K (0)$   & $0.780$ & $^{+0.022}_{-0.026}$ & $^{+0.008}_{-0.009}$& $^{+0.006}_{-0.010}$ & $^{+0.001}_{-0.000}$ \\
\hline
\end{tabular}
\end{table}

\begin{figure*}[tb]
\begin{center}
\includegraphics[width=0.34\textwidth]{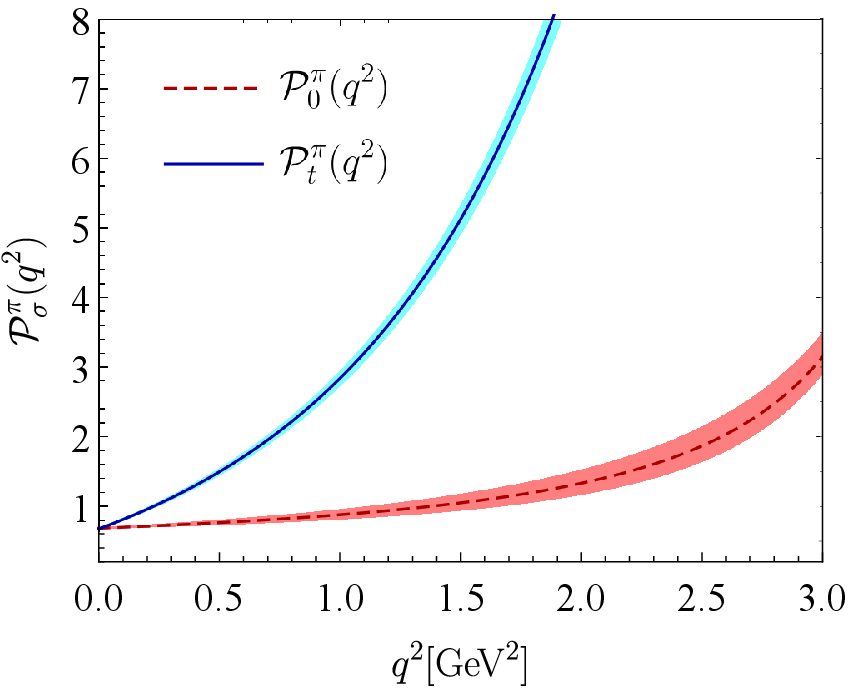}
\includegraphics[width=0.33\textwidth]{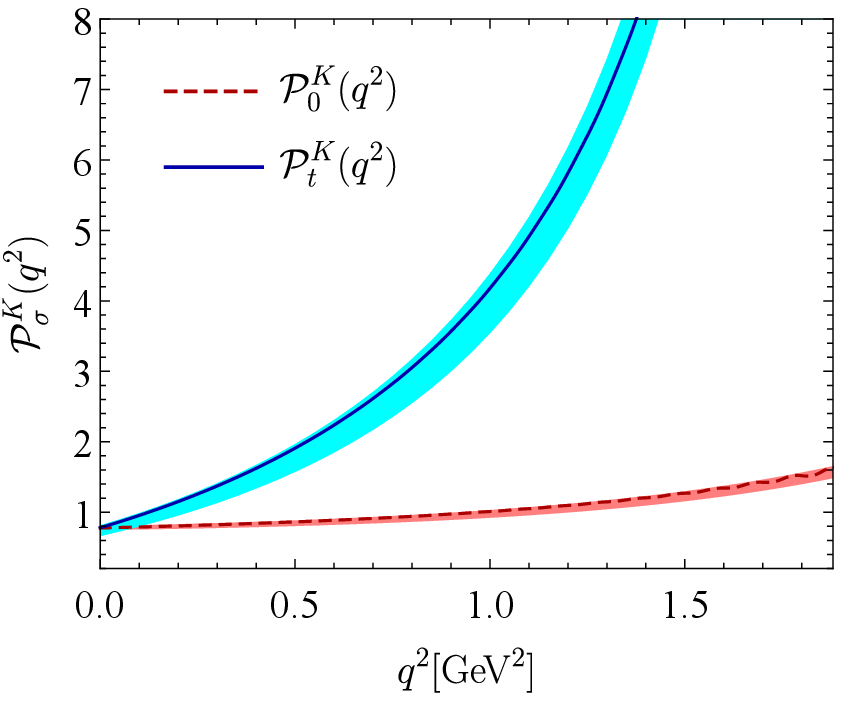}
\end{center}
\caption{The extrapolated LCSR predictions for the $D \to \pi (K)$ HFFs ${\cal P}_{0;t}^{P}(q^2)$, in which the shaded bands are squared average of those from the mentioned error sources.}
\label{HFF:P0t}
\end{figure*}
\begin{figure*}[!tb]
\begin{center}
\includegraphics[width=0.31\textwidth]{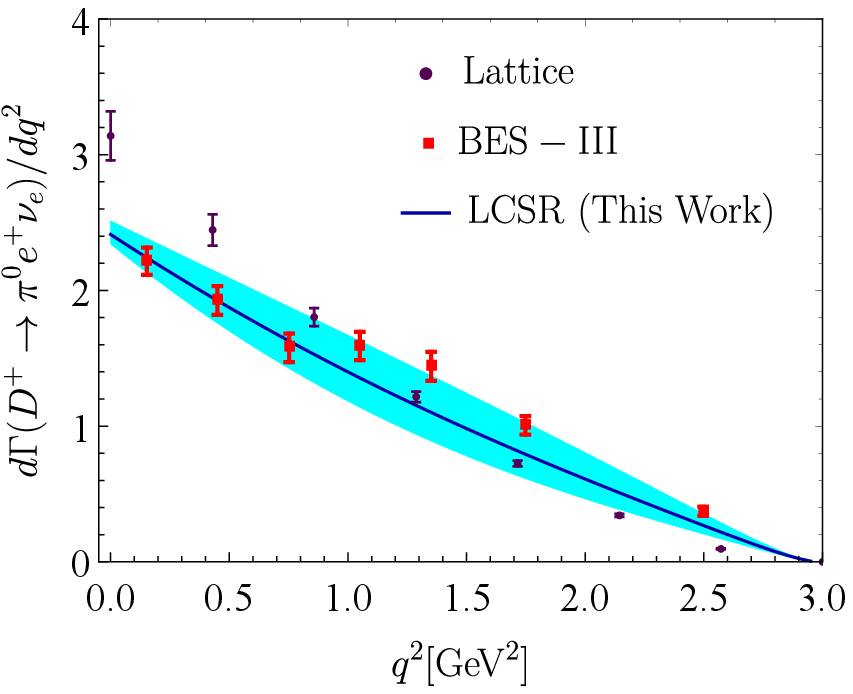}
\includegraphics[width=0.31\textwidth]{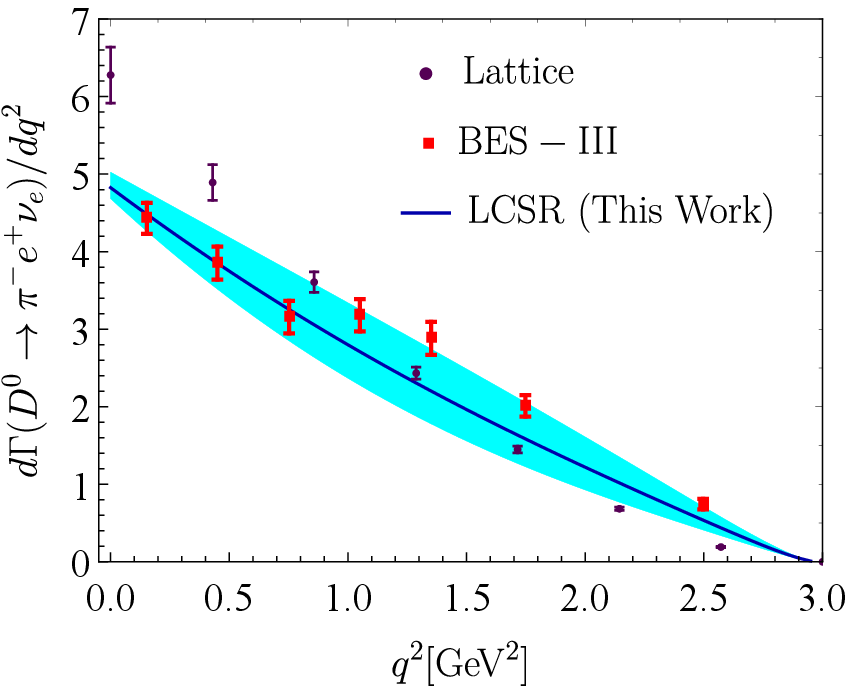}
\includegraphics[width=0.32\textwidth]{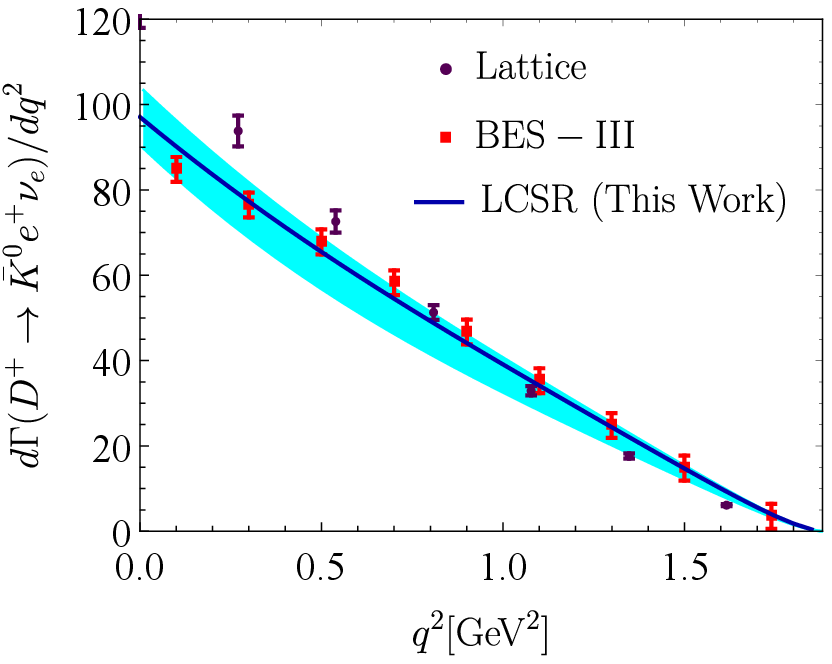}
\end{center}
\caption{The LCSR predictions for the $D\to P $ decay width, the solid lines represent center values and the shaded bands correspond to their uncertainties. The BES-III~\cite{ablikim:2017lks} results and Lattice~\cite{Lubicz:2017syv} results are also presented as a comparison.}
\label{fig:width}
\end{figure*}

\begin{table}[h]
\caption{The central value of $D\to P$ HFFs at the $q^2=0~{\rm GeV^2}$ for LO and NLO contributions, respectively.}
\begin{center}\centering
\begin{tabular}{c c c c }
\hline
~~$ $~~ &~~${\cal P}^K_{0;t}(0)$~~ &~~${\cal P}_{0;t}^\pi(0)$~~ \\
\hline
LO   & $0.757$ & $0.675$ \\
NLO  & $0.023$ & $0.013$ \\
Total & $0.780$& $0.688$ \\
\hline
\end{tabular}
\end{center}\label{tab:LONLO}
\end{table}
Theoretically, the LCSRs for $D\to P$ HFFs are applicable in low and intermediate $q^2$-regions, $q^2_{\rm LCSR, max} \simeq m_c^2-2m_cE$. Specifically, with the hadronic scale $E \approx 500 \rm{MeV}$~\cite{Khodjamirian:2000ds}, the $q^2_{\rm LCSR, max}$ can be taken as $0.6~\rm{GeV}^2$. While, the allowable physical range is $0 \leq q^2\leq q^2_{\rm max} = (m_D - m_P)^2$, i.e. $q^2\in [0, 3.0]{\rm GeV^2}$ for $D\to \pi$ and $q^2\in [0, 1.88]{\rm GeV^2}$ for $D\to K$, respectively. Based on the analyticity and unitarity properties of these HFFs, the extrapolation of these HFFs to the whole physical $q^2$-regions can be implemented via a rapidly converging series over the $z(t)$-expansion~\cite{Bharucha:2010im}. Specifically, the extrapolation of the HFFs satisfies the following parameterized formula,
\begin{align}
{\cal P}_{0}(t) &= \frac{1}{B(t) \, \phi_T^P(t)} \, \sum_{k=0}^{K-1} \alpha^{(0)}_k \, z^k(t) \,,
\cr
{\cal P}_{t}(t) &= \frac{1}{B(t) \, \sqrt{z(t,t_-)} \, \phi_L^P(t)} \, \sum_{k=0}^{K-1} \alpha^{(t)}_k \, z^k(t) \,,
\end{align}
where $\phi_L^P(t)=\phi_T^P(t)=1$, $\sqrt{z(t,t_-)}=\sqrt{\lambda}/m_D^2$ and
\begin{eqnarray}
z(t)=\frac{\sqrt{t_+ - t}-\sqrt{t_+ - t_0}}{\sqrt{t_+ - t}+\sqrt{t_+ - t_0}},
\end{eqnarray}
with $t_\pm=(m_D \pm m_P)^2$ and $t_0=t_+(1-\sqrt{1-t_-/t_+})$. The $B(t)=1- q^2/m_{R,i}^2$ is a simple pole corresponding to the first resonance in the spectrum. The resonance masses of quantum number $J^P$ are essential for the parameterisation of $D\to P$ HFFs ${\cal P}_{0;t}^{P}$ ~\cite{Tanabashi:2018oca, Leljak:2019eyw}, i.e.
\begin{align}
F_i = {\cal P}_0^P,~~J^P=1^-,~~m_{R,i}=2.010~{\rm GeV} \cr
F_i = {\cal P}_t^P,~~J^P=0^+,~~m_{R,i}=2.007~{\rm GeV} \nonumber
\end{align}
\begin{table}[t]
\centering
\caption{The fitted parameters $a_{0;1;2}^{P}$ for the HFFs ${\cal P}_{0;t}^{P}$, where all input parameters are set to be their central values.}\label{tab:SE}\centering
\begin{tabular}{c c c c c c}
\hline
~~&$~~{\cal P}_{0}^\pi$~~ & ~~${\cal P}_{t}^\pi$~~ & ~~${\cal P}_{0}^K$~~ & ~~${\cal P}_{t}^K$~~ \\ \hline
$a_0$    & 0.672    & $1.875$   & $0.762$   & $1.848$  \\
$a_1$    & $-0.354$ & $0.708$   & $-0.784$  & $-19.217$ \\
$a_2$    & 2.607   & $-47.245$ & $20.725$  & $-50.443$ \\
$\Delta$ & 0.00016  & $0.00022$ & $0.00758$ & $0.00000$ \\ \hline
\end{tabular}
\end{table}
The parameters $a_k^{\sigma}$ can be determined by requiring the ``quality'' of fit $\Delta < 0.01$, where $\Delta$ is defined as
\begin{equation}
\Delta=\frac{\sum_t\left|{\cal P}_{\sigma}(t)-{\cal P}_{\sigma}^{\rm fit}(t)\right|} {\sum_t\left|{\cal P}_{\sigma}(t)\right|}, \label{delta}
\end{equation}
where $t\in[0,0.06,\cdots,0.54,0.6]~{\rm GeV}^2$. The determined parameters $a_{k}^{\rho,\sigma}$ are listed in Table~\ref{tab:SE}, in which all the input parameters are set to be their central values. We put the extrapolated $D\to P$ HFFs ${\cal P}_{\sigma}^P(q^2)$ in Fig.~\ref{HFF:P0t}, where the shaded band stands for the squared average of all the mentioned uncertainties. All the HFFs increases with the increase of the squared momentum transfer.

\begin{table*}[htbp]
\begin{center}
\caption{Branching fractions of $D\to P\ell^+\nu_\ell$ (in unit $10^{-2}$). The errors are squared averages of all the mentioned error sources. As a comparison, we also present the prediction for various methods.}\label{tab:BF}
\begin{tabular}{c c c c c c c c}
\hline
Channel & LCSR(This Work)  & BES-III \cite{ablikim:2017lks} & CLEO-c \cite{Besson:2009} &BABAR~\cite{Lees:2015}& Belle~\cite{Widhalm:2006wz} & PDG \cite{Tanabashi:2018oca} & CQM \cite{Soni:2017eug}\\
\hline
$D^+\to\bar K^0e^+\nu_e$    & $8.547^{+0.445}_{-1.108}$ & 8.60(6)(15) & 8.83(10)(20) &-&-&-& 8.84\\
$D^+\to\bar K^0\mu^+\nu_\mu$& $8.435^{+0.437}_{-1.100}$	& 8.72(7)(18) &-&-&-&-& 8.60 \\
$D^+\to\pi^0e^+\nu_e$       & $0.328^{+0.056}_{-0.044}$ & 0.36(8)(5)  & 0.405(16)(9) &-&-&-& 0.309 \\
$D^+\to\pi^0\mu^+\nu_\mu$   & $0.325^{+0.056}_{-0.044}$	 &-&-&-&-&-& 0.303		\\
$D^0\to K^-e^+\nu_e$        & $3.370^{+0.176}_{-0.437}$	& 3.505(14)(33)&3.50(3)(4)&-& 3.45(7)(20)&3.538(33)&3.46 \\
$D^0\to K^-\mu^+\nu_\mu$ 	& $3.325^{+0.172}_{-0.433}$ & 3.505(14)(33)&-&-&-&3.33(13)&3.36\\
$D^0\to \pi^-e^+\nu_e$ 		& $0.258^{+0.044}_{-0.035}$ & 0.295(4)(3)&0.288(8)(3)&0.277(7)(9)&0.255(19)(16)&-&0.239\\
$D^0\to \pi^-\mu^+\nu_\mu$	& $0.256^{+0.044}_{-0.035}$ &-&-&-&-& 0.238(24) & 0.235 \\
\hline
\end{tabular}
\end{center}
\end{table*}

\subsection{The semi-leptonic decay processes of $D$-meson}\label{sec:App}
The differential decay rate for the process involving pseudoscalar mesons $D \to P$ is given by~\cite{Zhong:2018exo}
\begin{align}
& \frac{d\Gamma}{d q^2}(D \to P \ell\bar{\nu}_\ell) =  \frac{G_F^2|V_{cq}|^2}{24\pi^3 m_D^2} (1-2\delta_\ell)^2 |\vec{p}_P|\bigg[(1+\delta_\ell)
\nonumber \\
&  \quad\times m_D^2 |\vec{p}_P|^2 \big| {\cal P}_{0}^{P}(q^2)\big|^2  + \frac{3\lambda \delta_\ell}{4}\big|{\cal P}_{t}^{P}(q^2) \big|^2\bigg],\label{eq:width}
\end{align}
with $\delta_\ell = m_\ell^2/(2q^2)$, $|\vec{p}_P|$ is three-momentum of the pseudoscalar mesons in the $D$-meson rest frame, the fermi coupling constant ${G_F}=1.166\times 10^{-5} {\rm GeV^{-2}}$, $|V_{cs}| = 0.944$ and $|V_{cd}|= 0.2155$ \cite{Chen:2018zsn}.

After taking the resultant HFFs and other input parameters into Eq.~\eqref{eq:width}, we can obtain the differential decay widths of $D\to P\ell\nu_\ell$ in the entire kinematical range of squared momentum transfer, which is shown in Fig.~\ref{fig:width}. Where the solid lines represent center values and the shaded bands correspond to their uncertainties. The BES-III collaboration~\cite{ablikim:2017lks} and Lattice~\cite{Lubicz:2017syv} results are also presented as a comparison. As the suppression of the factor $|\vec{p}_{P}|= \sqrt{\lambda}/(2m_D)$ in the decay width formula for the $D\to P\ell\bar\nu_\ell$ semilepton decay, the differential decay width monotonously decreases with the increment of $q^2$, which is clearly illustrated in Fig.~\ref{fig:width}. More specifically, comparing with the decreasing rate of differential decay width for BES-III, our predictions are almost identical with it, so our results agree with the BES-III measurements within errors.

As a further step, the branching fractions of $D\to P \ell^+ \nu_\ell$ can be obtain by employing $\tau({D^0}) = 0.410(2)~{\rm ps}$ and $\tau({D^+}) = 1.040(7)~{\rm ps}$ and making integration with squared momentum transfer, the results are collected in Table~\ref{tab:BF}. The relevant experiment results BES-III~\cite{ablikim:2017lks}, CLEO-c~\cite{Besson:2009}, PDG~\cite{Tanabashi:2018oca}, Belle~\cite{Widhalm:2006wz}, BABAR~\cite{Lees:2015} and theory results the covariant quark model (CQM)~\cite{Soni:2017eug} are also presented as a comparison. Our results on the branching fractions for $D\to \pi \ell^+ \bar\nu_\ell$ and $D\to K \ell^+ \bar\nu_\ell$ are consistent with the BES-III measurements within errors. Compared with the CQM theoretical predictions for the $D^+\to \pi^0\ell^+\nu_\ell$ channel, our results are closer to the experimental results.

\begin{figure*}[!tb]
\begin{center}
\includegraphics[width=0.225\textwidth]{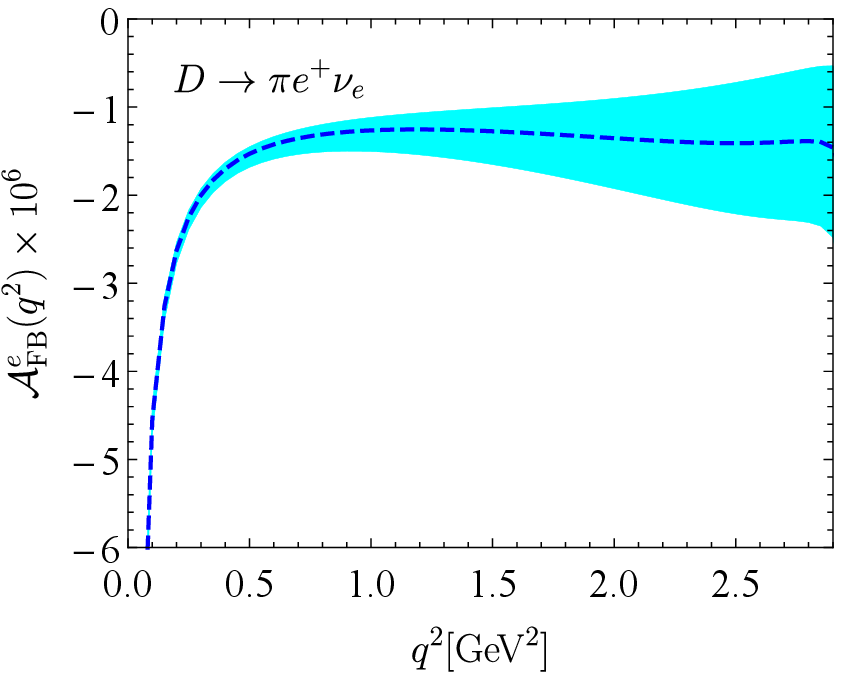}\includegraphics[width=0.238\textwidth]{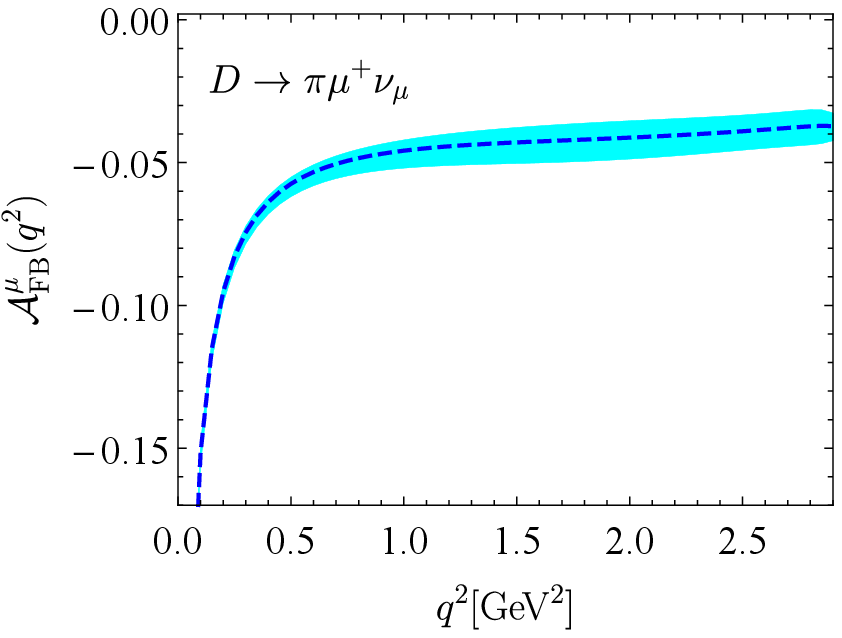}\includegraphics[width=0.257\textwidth]{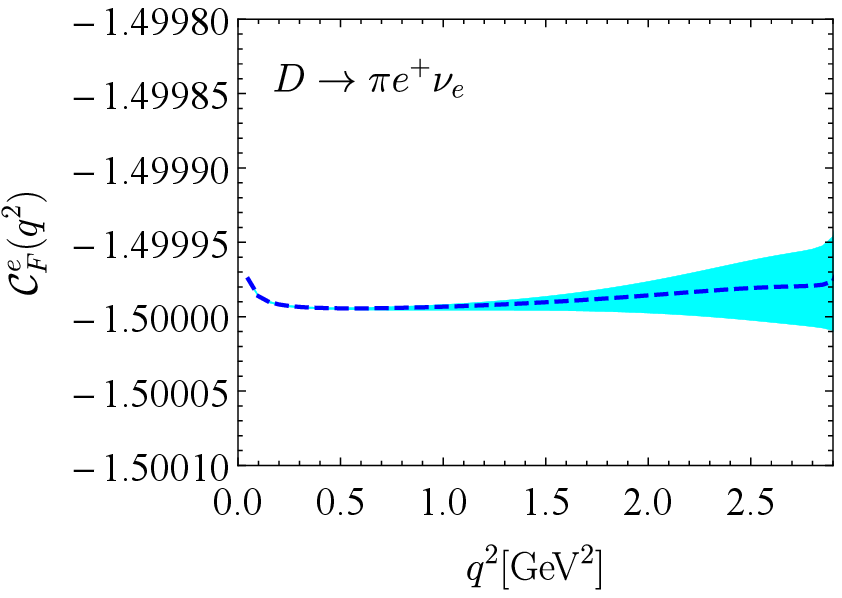}\includegraphics[width=0.233\textwidth]{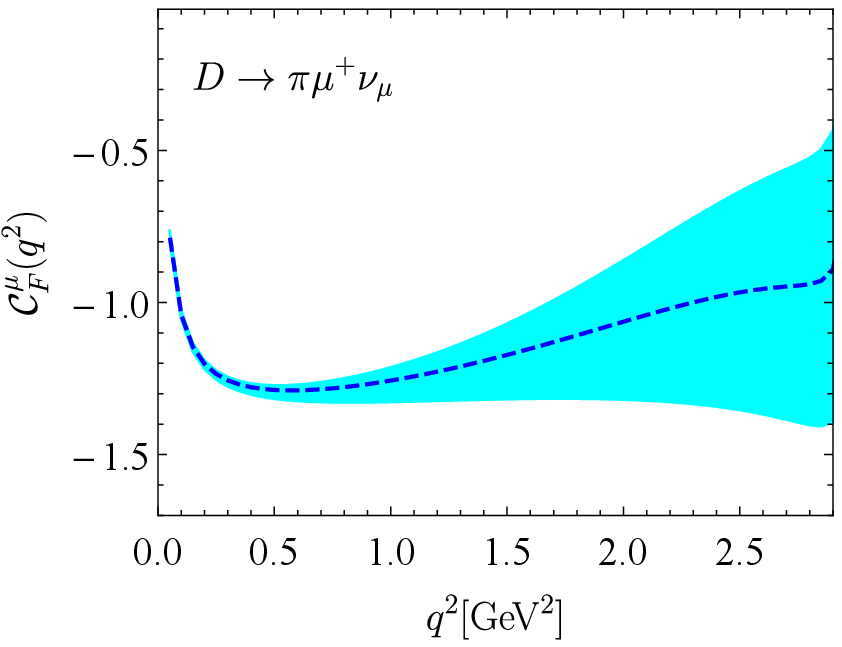}
\includegraphics[width=0.225\textwidth]{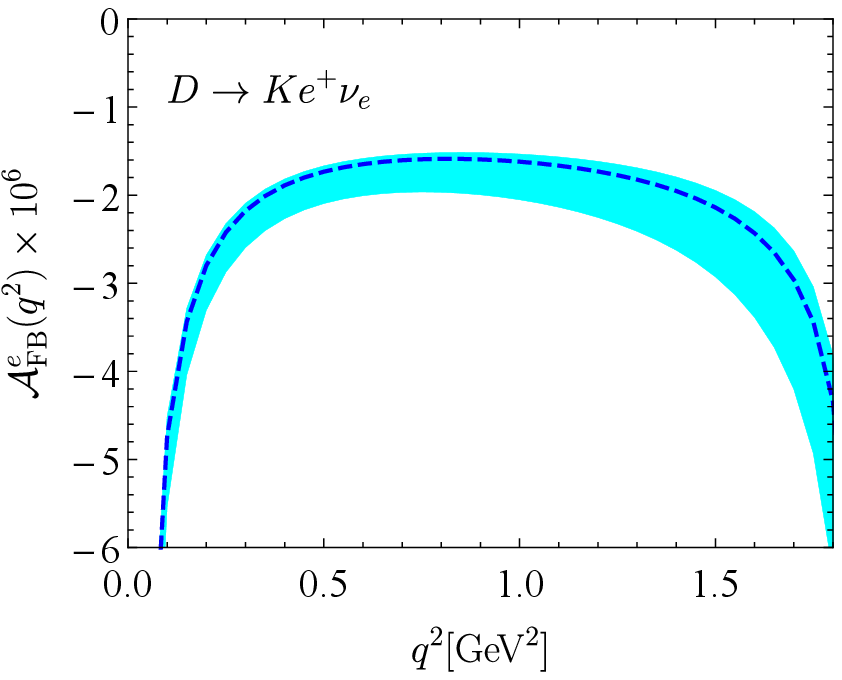}\includegraphics[width=0.238\textwidth]{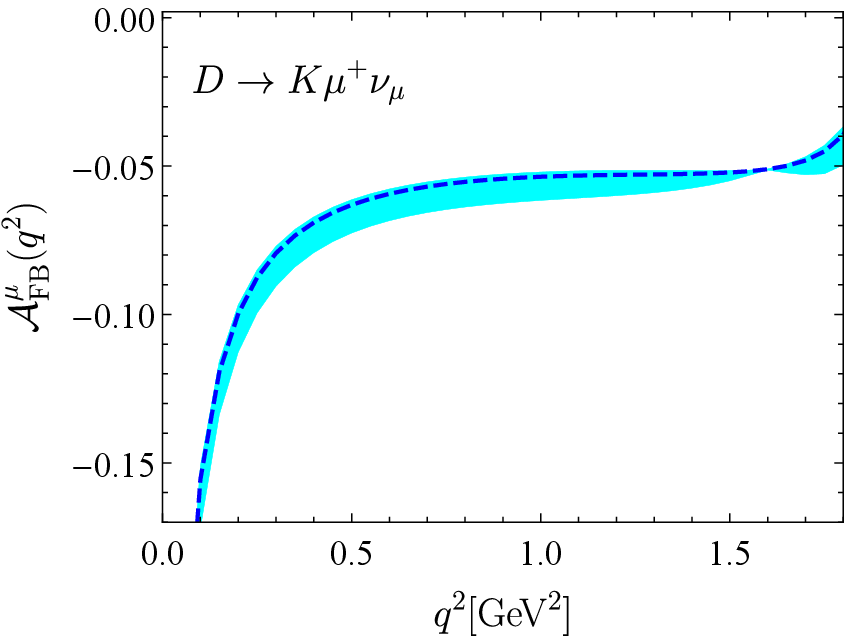}\includegraphics[width=0.257\textwidth]{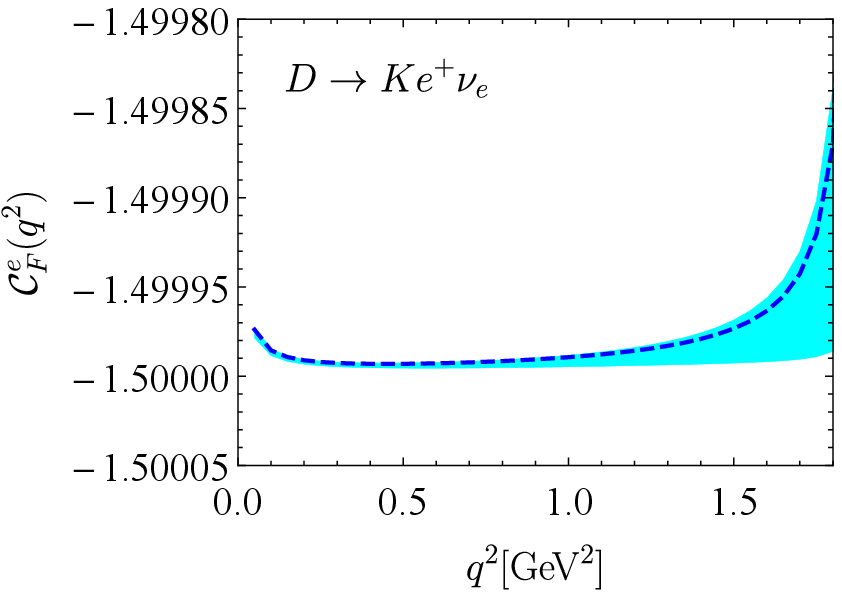}\includegraphics[width=0.233\textwidth]{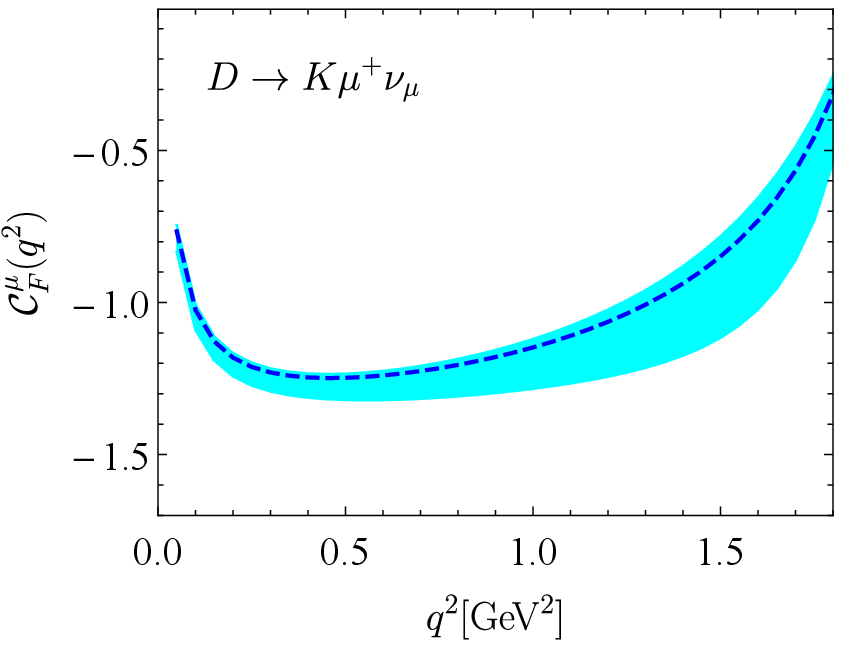}
\end{center}
\caption{The forward-backward asymmetries and the convexity parameters of the decays $D\to\pi\ell^+\nu_\ell$ and $D\to K\ell^+\nu_\ell$ with $\ell = e, \mu$, the shaded bands corresponds to the uncertainties from theoretical input parameters.}
\label{fig:AFBCFT}
\end{figure*}

Furthermore, the ratio for branching fractions for the different lepton channel are defined as,
$R_P = {\Gamma(D\to P\mu^+\nu_\mu)}/{\Gamma(D\to P e^+\nu_e)}.$
After taking the $D\to P\ell^+\nu_\ell$ decay width into the above equation and replacing HFFs with Eqs.~\eqref{eq:Av0} and \eqref{eq:Avt}, we can get ratios of semileptonic bractions as follows:

\begin{align}
R_{\pi^0} &= \frac{\Gamma(D^+ \to \pi^0  \mu^+\nu_\mu)}{\Gamma(D^+ \to \pi^0 e^+\nu_e)} = 0.991^{+0.230}_{-0.197}\nonumber\\
R_{\pi^-} &= \frac{\Gamma(D^0 \to \pi^- \mu^+\nu_\mu)}{\Gamma(D^0 \to \pi^- e^+\nu_e)} =
0.992^{+0.231}_{-0.198} \nonumber\\
R_{\bar K^0} &= \frac{\Gamma(D^+ \to \bar K^0 \mu^+\nu_\mu)}{\Gamma(D^+ \to \bar K^0 e^+\nu_e)} = 0.987^{+0.156}_{-0.138}\nonumber\\
R_{K^-} &= \frac{\Gamma(D^0 \to K^- \mu^+\nu_\mu)}{\Gamma(D^0 \to K^- e^+\nu_e)} = 0.987^{+0.156}_{-0.138}\label{eq:RatioBr}
\end{align}

The above ratios of our calculation are consistent with the CQM predictions $R_{\bar K^0}^{\rm CQM}= 0.97$~\cite{Soni:2017eug}, which are more close to $1$.

Moreover, the obtained HFFs can be used to calculate the forward-backward asymmetry ${\cal A}_{\rm FB}(q^2)$, which can be expressed as follows:
\begin{align}
{\cal A}_{\rm FB}^\ell(q^2) =  \frac{3\delta_\ell{\cal P}_0^P (q^2) {\cal P}_t^P(q^2)} {(1+\delta_\ell)|{\cal P}_0^P(q^2) |^2+3|{\cal P}_t^P(q^2) |^2} \label{eq:AFB}
\end{align}
Also, the lepton side convexity parameter ${\cal C}_F^\ell(q^2) $ can be written as:
\begin{align}
&{\cal C}_F^\ell(q^2) = -\frac32\frac{(1-2\delta_\ell)|{\cal P}_0^P(q^2) |^2}{(1+\delta_\ell)|{\cal P}_0^P(q^2) |^2+3\delta_\ell|{\cal P}_t^P(q^2) |^2}, \label{eq:CFl}
\end{align}
After taking the HFFs into Eqs.~\eqref{eq:AFB} and \eqref{eq:CFl}, we present the forward-backward asymmetry and the lepton convexity parameter within uncertainties in Fig.~\ref{fig:AFBCFT}. Finally, we list our the mean values for forward-backward asymmetry $\langle{\cal A}_{\rm FB}^\ell\rangle$, lepton convexity parameter $\langle{\cal C}_F^\ell\rangle$ in the Table~\ref{tab:AFBCFl}. It is seen that, for both $D\to \pi\ell\bar\nu_\ell$ and $D\to K\ell\bar\nu_\ell$ decay transitions, $\langle{\cal A}_{\rm FB}^\mu\rangle$ are about $10^5$ times larger than $\langle{\cal A}_{\rm FB}^e\rangle$, while $\langle{\cal C}_F^\mu\rangle$ is a little larger than $\langle{\cal C}_F^e\rangle$. For the $D\to P\ell\nu_\ell$, our predictions are different from the CQM results~\cite{Soni:2017eug}. Consider formula Eqs.~(\ref{eq:AFB}) and (\ref{eq:CFl}) refer to HFFs, these differences may provide a way to test those HFFs in future experiments.

\begin{table*}[t]
\centering
\caption{The forward-backward asymmetry and lepton convexity parameter. The errors are squared averages of all the mentioned error sources.}\label{tab:AFBCFl}
\begin{tabular}{c c c c c c c}
\hline
~Channel~& ~~$\langle{\cal A}_{\rm FB}^\ell\rangle$(This work)~~ & ~~$\langle{\cal C}_F^\ell\rangle$~(This work)~~ & ~~ $\langle{\cal A}_{\rm FB}^\ell\rangle$~(CQM)~\cite{Soni:2017eug}~~ & ~~$\langle{\cal C}_F^\ell\rangle$~(CQM)~\cite{Soni:2017eug}~~\\ \hline
$D\to\pi e^+\nu_e$    & $\big(-4.827^{+0.947}_{-1.247}\big)\times 10^{-6}$    & $-4.425^{+0.000}_{-0.000}$ & $-4.1\times 10^{-6}$ & $-1.5$ \\
$D\to\pi\mu^+\nu_\mu$ & $-0.155^{+0.013}_{-0.017}$                               & $-3.303^{+0.396}_{-0.509}$  & $-0.04$ & $-1.37$   \\
$D\to K e^+\nu_e$     & $\big(-4.564^{+0.316}_{-1.310}\big)\times 10^{-6}$    & $-2.775^{+0.000}_{-0.000}$  & $-4.27\times 10^{-6}$ & $-1.5$   \\
$D\to K \mu^+\nu_\mu$ & $-0.123^{+0.003}_{-0.015}$                              & $-1.866^{+0.063}_{-0.274}$ & $-0.058$ & $-1.32$ \\ \hline
\end{tabular}
\end{table*}

\section{Summary}\label{sec:summary}
In this paper, we investigate the $D \to P$ HFFs ${\cal P}_\sigma^P (q^2)$ with $\sigma = 0,t$ within LCSR approach up to NLO gluon radiation correction for twist-2 contributions accuracy. At large recoil point $q^2 \rightsquigarrow 0~{\rm GeV^2}$, we have ${\cal P}_{t,0}^\pi(0) = 0.688^{+0.020}_{-0.024}$, ${\cal P}_{t,0}^K(0)=0.780^{+0.024}_{-0.029}$. The detailed uncertainties of these predictions caused by the errors of the input parameters are given in Table~\ref{HFF uncertainties}. Then, contributions of the LO and NLO to the LCSR results are given in Table~\ref{tab:LONLO},and the maximal contribution of NLO for ${\cal P}_{0,t}^P(0)$ are no more than $3\%$. The results indicate those HFFs keep a high-accurate and the predictions we make with it are credible.

After extrapolating the $D\to P$ HFFs to whole physical $q^2$-region, the behavior of these HFFs within uncertainties are present in Fig~\ref{HFF:P0t}. Furthermore, we apply these HFFs to study the semilepton decays processes $D\to P\ell\nu_\ell$. For the decay width in Fig.~\ref{fig:width}, our results agree with BES-III collaboration within errors, especially in the low $q^2$ regions. With the help of $D$-meson lifetime, the two types of branching ratios are obtained and listed in Table~\ref{tab:BF}. Our predictions are in good agreement with the BES-III and other experimental results, which provides a better prediction of the $D\to \pi \ell\nu_\ell$ decay process than that of the CQM~\cite{Soni:2017eug}. Meanwhile, we list the ratio of branching fraction with different lepton channel $R_P$ in Eq.~\eqref{eq:RatioBr}, which shows that $R_{\bar K^0} = R_{K^-}$ and all values of these ratios close to 1.

As a further step, by taking the $D\to P$ HFFs into the forward-backward asymmetry ${\cal A}_{\rm FB}^\ell(q^2)$ and the lepton convexity parameter ${\cal C}_F^\ell(q^2)$, we give these two observable in Fig.~\ref{fig:AFBCFT}. Meanwhile, the mean values for the forward-backward asymmetry $\langle{\cal A}_{\rm FB}^\ell\rangle$ and lepton convexity parameter $\langle{\cal C}_F^\ell\rangle$ are listed in Table~\ref{tab:AFBCFl}. The table shows that $\langle{\cal A}_{\rm FB}^\mu\rangle$ are about $10^5$ times larger than $\langle{\cal A}_{\rm FB}^e\rangle$. There is a wide difference between our result and the CQM for both $\langle{\cal A}_{\rm FB}^\ell\rangle$ and $\langle{\cal C}_F^\ell\rangle$ for the $D\to P \ell\nu_\ell$ decay. The discrepancy may provide a way to test those HFFs in future experiments.
\\

\section*{Acknowledgments}
We are grateful to Prof. Xing-Gang Wu and Tao Zhong for many helpful discussions and suggestions. This work was supported in part by the National Science Foundation of China under Grant No.11765007, the National Natural Science Foundation of China under Grant
No.11947302, the Project of Guizhou Provincial Department of Science and Technology under Grant No.KY[2017]1089 and No.KY[2019]1171, the China Postdoctoral Science Foundation under Grant No.2019TQ0329.

\end{document}